\documentclass[11pt]{article}\textwidth 6.5in\textheight 9in
 \usepackage{amssymb} \usepackage [colorlinks] {hyperref}
\begin{document}

 \newcommand{\trm}[1]{{\bf\em #1}} \newcommand\emm[1]{{\ensuremath{#1}}}
 \newtheorem{thr} {Theorem} \newtheorem{lem} {Lemma}
 \newtheorem{prp} {Proposition} \newtheorem{cor} {Corollary}
 \newcommand\edf{{\stackrel{\mbox{\scriptsize df}}=}}

 \newcommand\m{\mbox{\bf m}} \newcommand\K{\mbox{\bf K}}
 \renewcommand\d{\mbox{\bf d}} \renewcommand\Pr{\mbox{\bf Pr}}
 \newcommand\I{\mbox{\bf I}} \newcommand\ov[1]{{\overline{#1}}}
 \newcommand\floor[1]{{\lfloor#1\rfloor}} \newcommand\ceil[1]{{\lceil#1\rceil}}
 \newcommand\lea{\prec}\newcommand\gea{\succ}\newcommand\eqa{\asymp}

 \newcommand\etc{{\em etc}} \newcommand\ie{{\em i.e.,\ }}
 \newcommand\eg{{\em e.g.,\ }} \newcommand\re{{\em r.e.\ }}
 \newcommand\RE{{\em R.e.\ }} \newcommand\M{\mbox{\bf M}}
 \newcommand\KM{\mbox{\bf KM}} \newcommand\E{\mbox{\bf E}}
 \newcommand\W{{\emm\Omega}} \renewcommand\a{{\emm\alpha}}
 \renewcommand\b{{\emm\beta}} \newcommand\g{{\emm\gamma}}
 \newcommand\w{{\emm\omega}} \newcommand\R{{\emm{\mathbb R}}}
 \renewcommand\r{{\emm\rho}} \newcommand\f{{\emm\varphi}}
 \renewcommand\l{{\emm\lambda}} \newcommand\Es{{\mbox{$\cal E$}}}

 \newcommand\II{{\makebox[0pt][l]{\hspace*{1pt}I}}}
 \newcommand\N{{\mbox{\II N}}} \newcommand\Q{{\mbox{\II Q}}}
 \newcommand\hreff[1]{{\small\url{http://#1}}}
 \newcommand\suf{{\mbox{\bf sf}}} \newcommand\qed{\rule{1em}{1.5ex} }

 \frenchspacing\title{\vspace*{-3pc} Forbidden Information}

\author {Leonid A.~Levin\\ Boston University\thanks
 {Computer Science department, 111 Cummington Mall, Boston, MA
02215.\newline This work (and updates in \hreff{arxiv.org/abs/cs/0203029} )
was supported by NSF grant CCF-1049505.\newline An earlier version written
for FOCS 2002 while the author worked at the Institut Des Hautes Etudes
Scientifiques. It was dedicated to the memory of Andrei Kolmogorov in the
100th year since his birth (4/25/1903).}}

\date{}\maketitle\begin{abstract} G\"odel Incompleteness Theorem leaves
open a way around it, vaguely perceived for a long time but not clearly
identified. (Thus, G\"odel believed informal arguments can answer any math
question.) Closing this loophole does not seem obvious and involves
Kolmogorov complexity. (This is unrelated to, well studied before,
complexity quantifications of the usual G\"odel effects.)
 I consider extensions $U$ of the universal partial recursive predicate
(or, say, Peano Arithmetic). I prove that any $U$ either leaves an
$n$-bit input (statement) unresolved or contains nearly all information
about the $n$-bit prefix of any \re real {\r} (which is $n$ bits for
some \r). I argue that creating significant information about a \trm
{specific} math sequence is impossible regardless of the methods used.
Similar problems and answers apply to other unsolvability results for
tasks allowing multiple solutions, \eg non-recursive tilings.
\end{abstract}

\section {Introduction.}

D.Hilbert asked if Peano Arithmetic (PA: consisting of logic and
algebraic axioms and an infinite family of Induction Axioms) can be
consistently extended to a complete theory. The question was somewhat
vague since an obvious answer was ``yes'': just add to PA axioms a
maximal consistent set, clearly existing albeit hard to find.\footnote
{I assume PA is consistent: a separate can of worms.}
 K.G\"odel formalized this question as existence, among such extensions,
of recursively enumerable (\trm {r.e.}) ones and gave it a negative
answer. Its mathematical essence is the absence of total recursive
extensions of universal partial recursive predicates (p.r.p.).
 This negative answer apparently was never accepted by Hilbert, and
G\"odel himself had reservations:\begin{quote}
 ``Namely, it turns out that in the systematic establishment of the
 axioms of mathematics, new axioms, which do not follow by formal logic
 from those previously established, again and again become evident.
 It is not at all excluded by the negative results mentioned
 earlier that nevertheless every clearly posed mathematical
 yes-or-no question is solvable in this way. For it is just this
 becoming evident of more and more new axioms on the basis of the
 meaning of the primitive notions that a machine cannot imitate.''
 \cite {goedel}\end{quote}

As is well known, \cite {brzd2,soare}, the absence of algorithmic solutions
is no obstacle when the task does not make a solution unique.\footnote
 {Note that no such problems arise for the less dramatic version of G\"odel
Theorem which makes the completion unique by requiring (unfalsifiable, thus
too abstract for Hilbert's program) \w-consistency, \ie that $\exists xA(x)
$ cannot be proven if $A(x)$ is refutable for each specific constant $x$.}
 A notable example is generating strings of linear Kolmogorov complexity,
\eg those that cannot be compressed to half their length. Algorithms fail,
but a set of dice does a perfect job! Thus, while \re sets of axioms cannot
complete PA, completion by other realistic means remained an open
possibility: one can so construct an \re theory $R$ that, like PA, allows
no consistent completion with \re axiom sets. Yet, $R$ allows a recursive
set of {\em pairs} of axioms such that random choice of one in each pair
assures such completion with probability 99\%. This cannot be done for PA
itself. In fact, \cite{fs} shows that any Martin-L\"of random sequence that
computes (i.e. allows computing from it) a consistent completion of PA also
computes the Halting Problem $H$; and by \cite {LMSS}, only a recursive
sequence (which $H$ is not) can be computed with a positive probability by
randomized algorithms.

Of course, G\"odel did not envision Math axioms to be chosen at random
:-).\\ {\bf But for arbitrary, not random, PA completions the reduction
  arguments do not work:} only a recursive predicate can be computed from
{\em all} consistent completions of PA.

However, the impossibility of a task can be formulated more generically.
\cite {K65} defined a concept of {\bf mutual information} in two finite
strings. It can be refined and extended to infinite sequences, so that it
satisfies conservation inequalities: cannot be increased by deterministic
algorithms or in random processes or with any combinations of both. In
fact, it seems reasonable to assume that no physically realizable process
can increase information about a specific sequence.

In this framework one can ask if non-mechanical means could really
enable the Hilbert-G\"odel task of consistent completion for PA
(as they can for the artificial system $R$ just mentioned).
 A negative answer follows from the existence of a {\em specific} sequence
{\r} that has infinite mutual information with {\em each} total extension
of a universal p.r.p. {\r} plays a role of a password: no substantial
information about it can be guessed, no matter what methods are allowed.

Note that invoking G\"odel's name, does not mean my intent to consider
widely discussed complexity aspects and implications of incompleteness
theorem. In particular, I ignore complexity of completions of PA. Much
of this was considered in the 60-s,\footnote
 {The original publications, such as \cite {brzd1}, gave the technical cores
of the results and avoided discussion of straightforward implications for
formal theories, G\"odel Theorem, etc. These implications were discussed
in talks, \eg mentioned in \cite {K71} with some technical references.
Also, extensive discussions were subsequently made by G. Chaitin.
 I omit details, since these results are only superficially related to the
 issues I address here.}
 but does not answer our question: are such completions really possible?
Strings of {\em any} complexity are easy to generate.

There are other interesting situations with a similar gap between the
proven result and its usual interpretation. Let me mention tiling, a cute
task studied in many areas of CS, Math, Physics, etc. A tile is a unit size
square with colored edges. A \trm {palette} is a finite set of tiles with
copies of which one can tile the plane so that adjacent edges match in
colors. Classical papers by Berger, Meyers, and others constructed palettes
$P$ that can tile an infinite plane, but only non-recursively, which is
typically interpreted as an impossibility of tiling. There, any program $t:
\N^2{\to}P$ can tile only finitely many frames $F=\{(i,j):\max(|i|,|j|){=}
n\}$, $\|F\|\edf n$ so that $t|_F$ appear on $P$-tiled planes. \cite{DLS}
 pushes these results to the limit, with a $P$ for which $\|F\|<\|t\|$.

Such palettes, thus, only allow tilings with frames of linear complexity.
This stronger result, though, makes the standard interpretation suspicious:
may these frames be just random, thus easy to generate with dice? Or could
more sophisticated and yet realistic means work?\\
 For some palettes this is, indeed, the case, but not for all. Like all
co-\re sets, the set of planar tilings with any given palette has members
with information about any specific sequence growing with radius $n$ as
slowly as $\log n$. Still, this bound cannot be improved for some palettes.

The same holds for complete extensions of universal {p.r.p.} and formal
systems. Thus, G\"odel Theorem is not really misleading. The proof of
 this (much trickier than G\"odel's) is the main point of this article.
 The interpretation of these results relies on {Independence Postulate} ---
an extended form of Church-Turing Thesis discussed in the last section.
 It is much stronger and can be applied where CT cannot.
 Other examples are results of \cite {L16} or older ones in \cite {L84}.

\newpage\section {Complexity Tools.}

\paragraph {Informal Overview.} A universal measure $\m$ is a largest up to
a constant factor enumerable by algorithms probability distribution. Its
entropy ($-\log$ of probability) is Kolmogorov Complexity $\K$; it is also
the least length of programs generating $x$. Rarity $\d(x|\mu)$ of $x$ for
a measure $\mu$ is $-\log\mu(x)$ minus complexity of $x$, \ie the
difference between entropies of $\mu$ and $\m$ for $x$. Mutual information
in $x,y$ is $\K(x){+}\K(y){-}\K(x,y)$: it allows to encode $x,y$ together
with fewer bits than separately. It is also their non-independence: rarity
for distribution $\m(x)\m(y)$ (universal on each, but independent on the
pair). It satisfies conservation inequalities: cannot be increased in
processing $x$, by deterministic algorithms, random transformations, any
their combinations, etc.

\paragraph {Conventions.} Let \R, \Q, \N, $B{=}\{0,1\}$, $S{=}B^*$,
$\W{=}B^\N$ be, respectively, the sets of non-negative reals, rationals,
integers, bits, finite, and infinite binary sequences; $x _{[n]}$ is the
$n$-bit prefix and $\|x\|$ is the bit-length of $x{\in}S$. A real function
$f$ and its values are \trm {enumerable} or \re ($-f$ is co-\re) if its
subgraph $\{(x,q): f(x)>q \in\Q\}$ is. \trm {Elementary} ($f{\in}\Es$) are
functions $f:\W\to\Q$ depending on a finite number of digits. $[A]\edf1$ if
a statement $A$ holds, else $[A]\edf0$. I identify objects (\eg integers)
with their binary encodings; $x{\in}S$ with $\Es$ functions $\w\mapsto
[x{=}\w_{[n]}]$ for $n{=}\|x\|$, etc. \trm {Majorant} is an \re\ function
largest, up to a constant factor, among \re functions in its class.
${\lea}f$, ${\gea}f$, and ${\eqa}f$ denote ${<}f{+}O(1)$, ${>}f{-}O(1)$,
${=}f{\pm}O(1)$, respectively.

\subsection {Integers: Complexity, Randomness.}

Let us define Kolmogorov \trm {complexity} $\K(x)$ as $\floor{1{-}\log\m
(x)}$ where $\m:\N\to\R$ is the \trm {universal measure}, \ie a majorant
\re function with $\sum_x\m(x){\le}1$. It was introduced in \cite{ZL},
and noted in \cite {L73,L74,Gacs74} to be a modification (restriction to
self-delimiting codes) of the least length of binary programs for $x$
defined in \cite {K65}. While technically different, {\m} relies on
intuition similar to that of \cite {Sol}. The proof of the existence of
the largest function was a straightforward modification of proofs in
\cite {Sol, K65} which have been a keystone of the informational
complexity theory.

For $x{\in}\N,y{\in}\N$ or $y{\in}\W$, similarly, $\m(\cdot|\cdot)$ is
the largest \re real function with $\sum_x\m(x|y){\le}1$; $\K(x|y)
\edf\floor{1{-}\log\m(x|y)}$ ($=$ the least length of self-delimiting
programs transforming $y$ into $x$).

\cite {K65} defines \trm {rarity} (non-randomness) $\d(x)$ of uniformly
distributed $x$ as $\|x\|{-}\K(x)$. Our modified {\K} allows extending this
to other measures $\mu$ on $\N$. A $\mu$-test is $f:\N\to\R$ with mean
$\mu(f){\le}1$ (and, thus, small values $f(x)$ on randomly chosen $x$).
 For computable $\mu$, a majorant \re test is $\m(x)/\mu(x)$.
 This suggests defining $\d(x|\mu)$ as $\ceil{\log\ceil{\m(x)/\mu(x)}}
\eqa\max\{0,\ceil{{-}\log\mu(x)}-\K(x)\}$.

\subsection {Integers: Information.}

In particular, $x{=}(a,b)$ distributed with $\mu{=}\m\otimes\m$, is a pair
of two independent, but otherwise completely generic, finite objects. Then,
$\I(a:b)\edf\d((a,b)|\m \otimes\m){=}\K(a){+}\K(b){-}\K (a,b)$ is seen as
\trm {dependence} and also measures \trm {mutual information} in $a,b$. It
was shown (see \cite {ZL}) by Kolmogorov and Levin to be close (within
${\pm}O(\log\K(a,b))$) to the expression $\K(a){-}\K(a|b)$ of \cite{K65}.
Our {\I} equals $\eqa\K(a)-\K(a|(b,\K(b)))$. Unlike the 1965 expression
(see \cite {Gacs74}), it is symmetric, monotone:
$\I(a:b)\lea\I((a,a'):b)$ (which will allow extending {\I} to $\W$),
and satisfies the following Independence Conservation inequalities \cite
{L74,L84}:
 For any computable transformation $A$, measure $\mu$, and some
family $t_{a,b}$ of $\mu$-tests \[\I(A(a):b)\lea \I(a:b),
\hspace{4pc} \I((a,w):b)\lea\I(a:b)+\log t_{a,b}(w).\]
 (The $O(1)$ error terms reflect the constant complexities of $A,\mu$.)
 So, independence of $a$ from $b$ is preserved in random processes,
 in deterministic computations, their combinations, etc. These
inequalities are not obvious (and false for the original 1965 expression
$\I(a:b){=}\K(a){-}\K(a/b)$~) even with $A$, say, simply cutting off
half of $a$. An unexpected aspect of $\I$ is that $x$ contains all
information about $k{=}\K(x)$, $\I(x:k)\eqa\K(k)$, despite
$\K(k|x)$ being ${\sim}\|k\|$ or ${\sim}\log\|x\|$, in the worst case
\cite{Gacs74}. One can view this as an ``Occam Razor'' effect: with no
initial information about it, $x$ is as hard to obtain as its simplest
($k$-bit) description.

All this works as well for the $\I_z$ variation of $\I$ allowing
all algorithms access to oracle $z$.

\subsection {Complexity, Randomness, and Information for Reals.}

We now extend these concepts to reals $\a{\in}\W$. This abstraction
is often convenient (if not taken too far) for concealing $O(1)$
terms and other small mismatches in formulas for finite objects.

\paragraph {Reals: Randomness.} A {measure} on $\W$ is a function $\mu
(x){=}\mu(x0){+}\mu(x1)$, for $x {\in}S$. Its mean $\mu(f)$ is a linear
functional on \Es: $\mu(f{+}g) {=}\mu(f) {+}\mu(g)$. It extends to other
functions, as usual. $\mu$-tests are lower semicontinuous $f$, $\mu(f){
\le}1$; computable $\mu$ have \trm {universal} (\ie majorant {\em r.e.})
\mbox{Martin-L\"of} tests $T_\mu(\a){=}\sum_i\m(\a_{[i]})/\mu(\a_{[i]})$.
\trm {Random} are $\a{\in}\W$ with rarity
$\d(\a|\mu)\edf\ceil{\log\ceil{T_\mu(\a)}}{<}\infty$.

\paragraph {Random \RE Reals.} P. Martin-L\"of noted that some random reals
are definable in arithmetic.\\ In fact, the least real $L{\in}[0,1]$ with
minimal rarity (see, \eg \cite {ZL}, section~4.4) is random and \re If
$(Y{+}Z)/X{\in}\Q$, we say $X$ \trm {dominates} \re reals $Y,Z$ (Solovay
reducibility, \cite {Slv,dh}). By \cite {KS}, random are exactly those \re
reals that dominate all others.\footnote
 {Indeed, let $X{=}L{+}Y$, $T$ be an \re test with $T(X){=}\infty$, $Y{=}
\sup_iy_i$, $L{=}\sup_il_i$, $T{=}\sum_it_i$; $l_i,y_i,t_i$ computable.
We can choose $T$ with $t_i(r)$ non-increasing for $r{\ge}l_i{+}y_i$.
Then $T'(r){\edf}\sum_it_i(r+y_i)$ is an \re test with $T'(L){=}\infty$.
 Conversely, let $Z,X\in[0,1]$ be enumerated as $Z{=}\sup_iz_i$,
$X{=}\sum_ix_i$. Let $s_{k,i}{=} \max\{z_i,s_{k,i{-}1}{+} x_i/2^k\}$ and
$t_k(V){\edf}2^k\sup_i[z_i{\le} V{<}s_{k,i}]$. Then $t_k$ and $T{=}\sum_k
t_k/k^2$ are \re tests. Let $y_i{=}s_{k,i}{-}(s_{k,i{-}1}{+}x_i/2^k)$,
and $Y{=}\sum_iy_i$. $S_k{=}\sup_is_{k,i}{=}X/2^k{+}Y$ dominates $X$.
If $Z{\ne}S_k$ for all $k$, then $z_i{\le}Z{<}s_{k,i}$ for some $i$, so
$t_k(Z){=}2^k$ and $T(Z){=}\infty$.}
 Any random \re {\r} is $\sum_n\m(n)$ for some universal \m: {\r}
dominates all such sums and replacing {\m} with $\m{+}x/i$ for an \re
$x$ concentrated in $0$, preserves universality. \cite {Chaitin}
contains probably the first published discussion of randomness of
$\sum_n\m(n)$.

\paragraph {Reals: Information.} \cite {L74} extends $\I$ to reals as
$\I(\a:\b)\edf\ceil{\log\ceil{\sum_{i,j}\m(i|\a)\m(j|\b)2^{\I(i:j)}}}$.\\
 (As always, we average in the linear scale and switch to the logarithmic
scale for the final expression.)

 For $\a,\b\in\N$, this equals our previous expression
 $\eqa\K(\a){+}\K(\b){-}\K(\a,\b)$ since both satisfy the
Independence Conservation Inequalities. In fact, this extension of $\I$
to $\W$ is the smallest satisfying the independence conservation.
 It suffices for the present paper and is used hereafter.

For other applications of Independence Postulate mentioned at the
end, a stronger (larger) expression can be used \cite {L84, L12}.
 It defines $\I(\a:\b)$ as $\d((\a,\b)|\M\otimes\M)$, where $\M$ is the
universal \re distribution on $\W$ (\ie\trm {semimeasure}: $\M(x){\ge}\M
(x0){+}\M(x1)$). This requires a (quite tricky) extension of the definition
of rarity $\d$ from computable measures to \re distributions.

\newpage\section {Consistent Objects.}

Consistency of theories and of other objects can be expressed as
membership in co-\re sets of reals. It is convenient to define such sets
via co-\re \trm {trees}, \ie infinite sets $T{\subset}S$ containing all
prefixes and {\bf some extensions of each} member. Let $\tilde T$ be the set
of those $\w\in\W$ with all prefixes in $T$. Some co-\re trees have only
strings of linear Kolmogorov complexity. Contrast this with
 \begin{prp}\label{log} For each $\b\in\W$, each co-\re tree $T$ has
$\a\in\tilde T$ with $\I(\a_{[n]}:\b)\lea 5\log n$.\end{prp}
 \begin{lem}\label{hint} For each co-\re tree $T$ there is a measure
$\mu(x){=}\mu (x0){+}\mu(x1)$ with $\mu (\tilde T)>1/2$,\\ computable as
$\mu(x){=}G(x,\r_{[5\log\|x\|]})$ by an algorithm $G$ using $5\log\|x\|$
digits of a \trm {hint} $\r\in\W$.\end{lem}
 Lemma~\ref {hint} implies Proposition~\ref {log}. Indeed, algorithms
can transform uniform distribution of inputs {\w} into any computable
one; same holds for computations with oracle \r. Consider an algorithm
using {\r} to compute $\mu$ and transforming {\w} into a
$\mu$-distributed {\a} ($\in\tilde T$ with high probability).
$\I(\r_{[O(\log\|x\|)]}:\b)= O(\log\|x\|)$; random {\w} cannot add
information with high probability, and the algorithm cannot increase it
either (due to conservation inequalities).
 \paragraph {Proof of Lemma~\ref{hint}:} $G$ uses {\r} to list all
converging $k$-bit programs. As \cite {brzd1} noted, for this it needs
just one of them, the slowest. {\r} can be any \re real with
$\K(\r_{[n]}){=} k{=}n{-}o(n)$, \eg a random one. Programs that use
$\r_{[n]}$ waiting for enumeration of \r's lower bounds to exceed
$\r_{[n]}$, are slower than any programs $P$ of complexity $<k{-}2\log
n$: otherwise $\r_{[n]}$ can be generated from $P,n$.

$G$ computes $\mu$ recursively in slices $\mu_i(x)$ for $\|x\|{=}n{=}
2^{2^i}$, assuming $\mu_{i{-}1}$ already computed.\\ It will approximate
$T\cap B^n$ as $T_i{=}T_i(\r)$ by limited co-enumeration and distribute
$\mu_{i{-}1}(x)$ uniformly on all $xy{\in}T_i$ (which always exist for
$x{\in}T$) or, if none, on $x0^*{\in}B^n$. Let $h_i$ be the Shannon entropy
of $\mu_i$, with the fractional part rounded up to $2\log 2i$ bits. Given
$\mu_{i{-}1}$, shrinking $T_i$ lowers $h_i$.

$G$ uses {\r} to compute the (lexicographically) least possible $h_1,
\ldots,h_i$ and co-enumerates $T_{\le i}$ until reaching these bounds.
Rounding $h_i$, leaves a fraction $f_i$ of $x\in T_i\setminus T$.
 Yet, $\sum_if_i<1/2$. \qed

In particular, randomized algorithms can generate strings of length $\ge
n$ of any co-\re tree $T$ with probability $1/k^2n, k{=}\ceil{\log n}$
by guessing $k,\floor{\log\|T\cap B^{2^k}\|}$.

\subsection {Example: Tiling}

An illustration is the tiling question from the introduction. \cite
{DLS} constructs a palette $P$ forcing, on each $P$-tiling, high
complexity of all horizontal tile strings not crossing one specific
column. The same construction works if complexity restriction is
replaced with membership in any \trm {bi-tree}, \ie a (co-{\em r.e.})
tree containing all substrings (not only prefixes) of its members. To
use it, we need to encode any co-\re tree $T$ as an equivalent bi-tree
$T_2$.

Let $b(2^k(2l{+}1))\edf$ $(l\bmod2)$. The pattern of $b(n)$ for $2^a$
consecutive $n$ determines the $a{-}2$ tail bits of $n$.
 Let $\ov i$ double each bit of $i$ and alter the result's first bit.
 If $n$ ends with $k$ followed by $\ov {\|k\|}$, let $\suf(n){\edf}k$.
Let $\tilde T_1$ be a tree of sequences $\a:\N{\to}B^2$ such that $\a(n)
{=}(b(n),t)$ and for some $s{\in}\tilde T$, whenever $\suf(n)$ is
defined, $t{=}s(\suf(n))$. Let $T_2$ be a bi-tree of all segments of
members of $\tilde T_1$. Each $n$-bit $T_2$-string represents the first
$n/O(\|n\|^2)$ bits\footnote
 {Optimizing $\ov i$ coding to $\|\ov i\|{=}\K(i)$ improves the
 overhead $\|n\|^2$ to $1/\m(\|n\|)$ but cannot eliminate it.}
 of a $T$-string $s$.

In particular, $T$ (and $T_2$) can force $s$ to be random \ie have maximal
complexity. This illustrates the point: all such tilings are highly
non-recursive, yet they are easy to generate (with dice). They can be
expressed in a formal system that allows trivial completions but no
recursive ones.

\newpage\section {The Taboo.}

This example does not show that all co-\re trees, such as tilings
with an arbitrary palette, allow easily generated members.
Proposition~\ref {log} sets a small but growing bound on the information
needed for that, leaving open the question the article started with. It
is resolved by the following observation central to this paper. We
represent in {\W} partial predicates as their graphs listed in arbitrary
order. Let $u$ be a universal partial recursive predicate (p.r.p.).

\begin{thr}\label{cens} Let $\r_{{\sim}n}$ be an $n{+}\K(n)$ bit prefix
of a random \re real {\r} and $U$ be a partial predicate that on $B^n$
is a total extension\footnote
 {One can weaken this total extension condition to being consistent with
$u$ and defined on the specific input $px_n$ if the simple $\I$ of \cite
{L74} is strengthened to one of \cite {L84} and $\M(P(0^{\|x\|})\ldots
P(x{-}1))$ replaces $m_{x,i}$ in $P$ of the proof.}
 $U_n$ of $u$. Then $\I(U:\r_{{\sim}n})\gea n-\K(\K(n)|n)$. \end{thr}

This statement that $U_n$ carries almost all information on $\r_n$
(or equivalently, on the domain size $d_n$ of $u|_{B^n}$ ) may
seem paradoxical. Indeed, Andrei Muchnik noted that only a recursive
sequence can be computed from all total extensions $U$ of $u$.
Laurent Bienvenu noted that some such $U$ have $\K(\r_n|U)\sim n$.
The explanation is that unlocking information in $U_n$ to compute $\r_n$
requires also knowing $\K(U_n)$. \cite {Gacs74} ingeniously proves that
$\K(\K(x)|x)$ can be $\sim\log(\|x\|)$, \ie $\sim n$ for $x=U_n$.
 Theorem~\ref{cens} provides an alternative proof of this.

Muchnik's observation does not apply to sequences with computable
complexities: indeed by \cite{fs}, any Martin-L\"of {\bf random} sequence
that computes a total extension of $u$, also computes the domain of $u$. Of
course, I do not assume the axioms chosen by math community to be random!

\paragraph {Proof.} We define a p.r.p.~$P:S{\to}B$ inductively on $B^n$.
If $P(x)$ is defined on $[0^n,x{-}1]$, let $\M_{x,i}$ denote the combined
universal measure $\sum_Q\m(Q|n)$ of all total predicates $Q$ on $B^n$
that agree with $P$ on $[0^n,x{-}1]$ and $Q(x){\ne}i{\in}B$.
 Then $P(x)$ enumerates lower bounds for $\M_{x,i}$ until either exceeds
$2^{-n}$ and yields $P(x){=}i$, decreasing $\sum_i \M_{x{+}1,i}$ by ${>}2
^{-n}$. For some $x_n$, $\M_{x_n,i}{\le}2^{-n}$ and $P$ diverges on $[x_n,
1^n]$ with $\sum_Q\m(Q|n){\le}2/2^n$ for all total extensions $Q$ of $P$
on $B^n$.

For all such $Q$, this bound allows $2^n/O(1)$-fold increase of
$\m(Q|x_n)$,\footnote
 {This also means $n{\eqa}I_n(x_n:Q){\lea}I_n(d_n:Q)$ (and already
 implies Corollary~\ref {pr}, even simplifies its bound to $O(2^{-n})$ ).
 So, $\K(d_n|(Q,\K(Q|n))\eqa0$ thus $d_n$ can be computed from
 $U_n,\K(U_n|n)$, if not from $U_n$ itself.}
  compared to just $\m(Q|n)=\m(\K(n)|n)\m(Q)/O(\m(n))$.
Now, $u(px)$, with a fixed $p$, computes $P(x)$, and $U(px)$ extends
$P(x)$ on $B_n$ to a total $Q_n$, with $\m(Q_n|U)=\m(n)/O(1)$.
Also, $x_n,\r_{{\sim}n},\K(n)=\|\r _{{\sim}n}\|{-}n$ are {\em r.e.},
so can be computed from one of them whose enumeration ends latest.
This could only be $\r_{{\sim}n}$, being random and long enough
to dominate in complexity (which computations cannot increase).

Thus, $\m(Q_n|\r_{{\sim}n})=\m(Q_n|x_n)/O(1)=2^n\m(\K(n)|n)\m(Q_n)/O(\m
(n))$. Then, $\I(U:\r_{{\sim}n})\ge$ $\log(\m(Q_n|\r_{{\sim}n})\m(Q_n|U)/\m
(Q_n))\gea \log2^n\m(\K(n)|n)\gea n{-}\K(\K(n)|n)$. \qed

Since random strings contain $k$ bits of information about {\r} only
with probability $2^{-k}$ and algorithms do not increase information
(due to the Conservation Inequalities), Theorem~\ref {cens} implies
 \begin{cor}\label{pr} The probability that a randomized algorithm computes
 on $B^n$ a total extension\\ of $u$ is at most $O(2^{-n})/\m(\K(n)|n)$.
 {\em (Strengthening the $o(1)$ bound of \cite {soare}.)} \end{cor}
 (Thus, not all palettes, formal theories, etc.
 allow randomness-based tilings, completions, etc.)

While nobody envisioned choosing fundamental Math axioms by coin flips,
Theorem~\ref{cens} supports a more general impossibility.
 Just like the usual interpretation of G\"odel Theorem is a matter
 of accepting Church-Turing Thesis, judging if Theorem~\ref {cens}
 makes the completion task impossible is a matter of accepting
 the Independence Postulate discussed below.

\vspace{-4pt}\section {The Independence Postulate.}\vspace{-6pt}

\paragraph {IP:} {\em Let $X$ be a sequence defined with an $n$-bit
mathematical statement (\eg in PA or set theory),
 and a sequence $Y$ can be located in the physical world with a $k$-bit
instruction set (\eg ip-address). Then $\I(X:Y)<k{+}n{+}c $, for some
small absolute constant $c$.}

(Note that $X$ and $Y$ can each have much more than $k{+}n{+}c$ bits of
information.)

Thus, a (physical) sequence of all mathematical publications has little
information about the (mathematical) sequence of all true statements of
arithmetic. This is of little concern because the latter has, in turn,
little information about the stock market (a physical sequence). :-)

Of course, Kolmogorov information is not the only desirable commodity.
Yet, IP has interesting applications \cite {L84}. It can be restated as
a ``finitary'' version of the Church-Turing thesis (CT) by calling \trm
{recursive} those finite sequences with recursive descriptions nearly as
short as any their ``higher-level'' math descriptions.
 IP postulates that only such recursive sequences exist in reality.

Let me add (in order of increasing relevance) some comparisons between
 IP and CT:\begin{enumerate}

\item IP is stated with greater care than CT: Obviously not all strings we
generate are algorithmic (non-communist election results better not be :-).
Only mathematically defined strings need be algorithmic to be
generatable. IP includes this math clause explicitly, CT rarely does.
 \item IP is simpler, CT more abstract. All sequences we ever see are
computable just by being finite: CT is useless for them!
 IP works equally well for finite and infinite sequences.
 \item IP is easier to support: CT is usually stated with vague
reasoning. IP has broad conservation laws to support it and a general
intuition that target information cannot be increased.
 \item IP is much more comprehensive: CT prohibits only generating the
target math sequence itself; IP bars all strings with any significant
information about it. So, IP can be applied where CT cannot. See, \eg
\cite {L84} or more recent results in \cite {L16}.\end{enumerate}

One application is dousing G\"odel's hope cited in the Introduction,
regardless of any realizable process of axiom selection.
 The argument is ``inductive''. It seems, complicated processes we
observe, can ultimately be {\em explained}, \ie reduced to simpler ones.
These reductions use deterministic models and random ones, but neither
can {\bf\em increase} the starting information about a target.
 The toolkit of our models may change (\eg quantum amplitudes work
somewhat differently than probabilities) but it is hard to expect
new realistic primitives allowing such ``information leaks''.

 So, if {\em complicated} processes generate unlimited target
information, so must do some {\em elementary} processes,
 that admit no further explanations (reductions to simpler processes).
The existence of such elementary unexplainable information Sources
cannot be ruled out. Yet Infidels :-) can postulate it away.
 Just like the impossibility of generating power from uniform heat,
 this is an unprovable postulate, supported by proven arguments.

Note that the above argument is based on Independence Conservation
Inequalities (ICI) of \cite{L74,L84}. They deal with generation of
strings by deterministic algorithms or by random processes {\bf\em from
other strings}. If the preexisting string has no significant target
information, neither will the generated one. And despite being
intuitive, ICI are not technically trivial and should {\bf not be
confused with the easy remark} that randomized algorithms cannot
generate {\bf\em from scratch} information about math targets, such as
\eg \re reals. (Math community never tried choosing their fundamental
axioms this way :-). But the difficulty pays off, being essential for
the inductive nature of the support ICI give to IP.

\section* {Acknowledgments} I am grateful to Robert Solovay,
Alexander Shen, Bruno Durand, and Laurent Bienvenu for insightful
discussions, and to Rod Downey, Denis Hirschfeldt, and Stephen Simpson
for three references.


\begin{thebibliography}{99}\setlength{\itemsep}{0pt}\footnotesize

\bibitem [DAN]{DAN} {\em Doklady AN SSSR = Soviet Math. Doklady}.

\bibitem [Barzdin 68] {brzd1} Janis M. Barzdin. Complexity of programs
to determine whether natural numbers\\ not greater than $n$ belong
to a recursively enumerable set. \cite {DAN}, 9:1251-1254, 1968.

\bibitem [Barzdin 69] {brzd2} Janis M. Barzdin'.
O Vychislimosti na Verojatnostnyh Mashinah. \cite {DAN} 189(4), 1969.

\bibitem [Berger 66] {berger} R. Berger.
The undecidability of the domino problem. {\em Memoirs of AMS}, 66, 1966.

\bibitem [Chaitin 75] {Chaitin} G.J. Chaitin. A Theory of Program Size
Formally Identical to Information Theory.\\ JACM, 22:329-340, 1975.

\bibitem [dLMSS 56] {LMSS} Karel De Leeuw, Edward F. Moore, Claude E. Shannon,
N. Shapiro. Computability by\\ probabilistic machines. In {\em Automata
Studies}, Ann. Math. Studies 34. Princeton U. Press, 1956. 

\bibitem [Downey, Hirschfeldt 10] {dh} Rodney G. Downey, Denis R. Hirschfeldt.\\
 {\em Algorithmic Randomness and Complexity.} Springer, 2010.

\bibitem [Durand, Levin, Shen 01] {DLS} Bruno Durand, Leonid A. Levin,
Alexander Shen. Complex Tilings.\\ {\em J.Symb.Logic} 73/2:593-613, 2008.
Also: STOC 2001. \hreff {arXiv.org/abs/cs.CC/0107008}

\bibitem [Gacs 74] {Gacs74} Peter Gacs.
On the Symmetry of Algorithmic Information. \cite {DAN} 15:1477, 1974.

\bibitem [G\"odel 61] {goedel} Kurt G\"odel. The modern development
of the foundations of mathematics in the light of philosophy.\\ 1961.
In: Kurt G\"odel. Collected Works. Volume III. Oxford University Press.\\
\hreff {evans-experientialism.freewebspace.com/godel.htm}

\bibitem [Jockusch, Soare 72] {soare} Carl G. Jockusch, Jr.,
Robert I. Soare. $\Pi^0_1$ Classes and Degrees of Theories.\\
Trans. Am. Math. Soc. 173:33--56 1972.

\bibitem [Kolmogorov 65] {K65} Andrei N. Kolmogorov.
Three Approaches to the Concept of the Amount of Information.\\
{\em Probl. Inf. Transm.}, 1(1):1-7, 1965.

\bibitem [Kolmogorov 72] {K71} Andrei N. Kolmogorov. Complexity of
specifying and complexity of constructing\\ mathematical objects.
 Part 4. Talk at the 11/23 Meeting of Moscow Math. Society, 1971.\\
Abstract in {\em Uspekhy Mat. Nauk} 27(2), 1972. (In Russian.)

\bibitem [Kucera, Slaman 01] {KS} A. Kucera, T.A. Slaman. Randomness
and recursive enumerability.\\ SIAM J. Comput. 31:199-211, 2001.

\bibitem [L 73] {L73} Leonid A. Levin.
On the Concept of a Random Sequence. \cite {DAN} 14(5):1413-1416, 1973.

\bibitem [L 74] {L74} Leonid A. Levin. Laws of Information Conservation
(Non-growth) and Aspects of the Foundations\\ of Probability Theory.
{\em Probl. Pered. Inf.= Probl. Inf. Transm.} 10(3):206-210, 1974.

\bibitem [L 84] {L84} Leonid A. Levin. Randomness Conservation Inequalities.
 {\em Inf. \& Control} 61(1):15-37, 1984.

\bibitem [L 12] {L12} Leonid A. Levin. Enumerable distributions,
    randomness, dependence. 2012,\\ \hreff {arxiv.org/abs/1208.2955}

\bibitem [L 16] {L16} Leonid A. Levin. Occam Bound on Lowest Complexity
 of Elements. APAL 167:958-961, 2016.\\ \hreff {arxiv.org/abs/1403.4539}

\bibitem [Myers 74] {myers} D. Myers. Nonrecursive tilings of the plane.
ii. {\em J. Symb. Logic}, 39(2):286--294, 1974.

\bibitem [Solomonoff 64] {Sol} R.J. Solomonoff.
A Formal Theory of Inductive Inference. {\em Inf. \& Cntr} 7(1), 1964.

\bibitem [Solovay 75] {Slv} R. Solovay. Unpublished manuscript.
 IBM Watson Res. Ctr, New York 1975.

\bibitem [Stephan 06]{fs} Frank Stephan. Martin-L\"of random and
PA-complete sets. {\em Lecture Notes in Logic,} 27:342--348, 2006.

\bibitem [ZL 70] {ZL} Alexander Zvonkin, Leonid A. Levin. The complexity of
finite objects and the algorithmic\\ concepts of information and randomness.
{\em UMN = Russian Math. Surveys} 25(6):83-124, 1970.

\end{thebibliography}
 \end{document}